\documentclass[
    journal=largetwo,
    manuscript=preprint,
    year=2025,
]{cup-journal}

\usepackage{amsmath}
\usepackage[nopatch]{microtype}
\usepackage{booktabs}
\usepackage[export]{adjustbox}
\usepackage{float}
\usepackage{graphicx}
\usepackage{array}
\usepackage{tabularx}
\usepackage{multirow}

\title{Novel AI-Based Quantification of Breast Arterial Calcification to Predict Cardiovascular Risk}

\author{Theodorus Dapamede}
\shortauthor{Dapamede}
\affiliation{Department of Radiology, Emory University, Atlanta, GA, USA}
\author{Aisha Urooj}
\affiliation{Department of Radiology, Mayo Clinic, Phoenix, AZ, USA}
\author{Vedant Joshi}
\affiliation{Department of Radiology, Mayo Clinic, Phoenix, AZ, USA}
\author{Gabrielle Gershon}
\affiliation{Department of Radiology, Emory University, Atlanta, GA, USA}
\author{Frank Li}
\affiliation{Department of Radiology, Emory University, Atlanta, GA, USA}
\author{Mohammadreza Chavoshi}
\affiliation{Department of Radiology, Emory University, Atlanta, GA, USA}
\author{Beatrice Brown-Mulry}
\affiliation{Department of Radiology, Emory University, Atlanta, GA, USA}
\author{Rohan Satya Isaac}
\affiliation{Department of Radiology, Emory University, Atlanta, GA, USA}
\author{Aawez Mansuri}
\affiliation{Department of Radiology, Emory University, Atlanta, GA, USA}
\author{Chad Robichaux}
\affiliation{Department of Biomedical Informatics, Emory University, Atlanta, GA, USA}
\author{Chadi Ayoub}
\affiliation{Department of Cardiovascular Medicine, Mayo Clinic, Phoenix, AZ}
\author{Reza Arsanjani}
\affiliation{Department of Cardiovascular Medicine, Mayo Clinic, Phoenix, AZ}
\author{Laurence Sperling}
\affiliation{Department of Medicine, Emory University, Atlanta, GA}
\author{Judy Gichoya}
\affiliation{Department of Radiology, Emory University, Atlanta, GA, USA}
\author{Marly van Assen}
\affiliation{Department of Radiology, Emory University, Atlanta, GA, USA}
\author{Charles W. O’Neill}
\affiliation{Department of Medicine, Emory University, Atlanta, GA}
\author{Imon Banerjee}
\affiliation{Department of Radiology, Mayo Clinic, Phoenix, AZ, USA}
\author{Hari Trivedi}
\affiliation{Department of Radiology, Emory University, Atlanta, GA, USA}


\begin{document}

\begin{abstract}
IMPORTANCE
Women are underdiagnosed and undertreated for cardiovascular disease. Automatic quantification of breast arterial calcification on screening mammography can identify women at risk for cardiovascular disease and enable earlier treatment and management of disease. 

OBJECTIVE
To determine whether artificial-intelligence based automatic quantification of BAC from screening mammograms predicts cardiovascular disease and mortality in a large, racially diverse, multi-institutional population, both independently and beyond traditional risk factors and ASCVD scores.

DESIGN, SETTING, AND PARTICIPANTS
Retrospective cohort study of 116,135 women from two healthcare systems (Emory Healthcare and Mayo Clinic Enterprise) who had screening mammograms and either experienced a major adverse cardiovascular event, death, or had at least 5 years of clinical follow-up. BAC was quantified using a novel transformer-based neural network architecture for semantic segmentation. BAC severity was categorized into four groups (no BAC, mild, moderate, and severe), with outcomes assessed using Kaplan-Meier analysis and Cox proportional-hazards models.

MAIN OUTCOMES AND MEASURES
Major Adverse Cardiovascular Events (MACE), including acute myocardial infarction, stroke, heart failure, and all-cause mortality, adjusted for traditional risk factors and Atherosclerotic CVD (ASCVD) risk scores.

RESULTS
BAC severity was independently associated with MACE after adjusting for cardiovascular risk factors, with increasing hazard ratios from mild (HR 1.18-1.22), moderate (HR 1.38-1.47), and severe BAC (HR 2.03-2.22) across both internal and external datasets (all p<0.001). This association remained significant across all age groups, with even mild BAC indicating increased risk in women under 50 years. When analyzed in patients with ASCVD risk scores available, BAC remained an independent predictor of MACE, showing significant associations with acute myocardial infarction (HR 1.18-1.22), stroke (HR 1.10-1.23), heart failure (HR 1.22-1.27), and all-cause mortality (HR 1.17-1.30) (all p<0.005).

CONCLUSIONS AND RELEVANCE
Automated BAC quantification enables opportunistic cardiovascular risk assessment during routine mammography screening without additional radiation exposure or cost. This approach provides predictive value independent of traditional risk factors and ASCVD scores, particularly in younger women, suggesting potential value for early CVD risk stratification in the millions of women already undergoing annual mammography.

\end{abstract}

\section{Introduction}
Cardiovascular disease (CVD) is the leading cause of death in women yet is frequently underdiagnosed \cite{mikhail_coronary_2005, wenger_coronary_2003, kelsey_results_1993}. 
Risk assessment for CVD is typically performed using a combination of laboratory markers, family history, and other risk factors that are combined in various risk prediction models, with the Atherosclerotic CVD (ASCVD) score currently recommended by guidelines \cite{goff_2013_2014, score2_working_group_and_esc_cardiovascular_risk_collaboration_score2_2021, khan_development_2024}. 
For intermediate risk patients (ASCVD risk $\geq7.5\%$ to $<20\%$) and selected patients at borderline risk (5\% to <7.5\%), gated non-contrast computed tomography (CT) examinations can quantify the amount of coronary artery calcium (CAC) for estimation of coronary atherosclerotic burden and incremental prognostication \cite{arnett_2019_2019}. 
However, CAC scans involve additional cost and radiation exposure and therefore are not amenable to screening of large populations. In contrast, breast arterial calcification (BAC) is visible on routine screening mammography which is obtained in approximately 40 million women annually in the US and can be automatically quantified using Artificial Intelligence (AI) without additional time, cost, or radiation exposure.
\subsection{BAC as a marker of cardiovascular disease}
Multiple prior studies have shown that BAC correlates with calcification in peripheral arteries \cite{duhn_breast_2011} and can predict CVD events in the general population independent of traditional risk factors \cite{kemmeren_arterial_1998, iribarren_breast_2004}. Although most prior studies have considered only the presence or absence of BAC as a binary variable, BAC can be reliably quantified \cite{manzoor_progression_2018} with sufficient sensitivity to detect progression of disease and quantify underlying risk factors \cite{alappan_warfarin_2020, alappan_vascular_2020}. Studies have shown that although BAC predicted CVD, it has poor correlation with CAC scores on chest CT, further demonstrating its potential as an independent prognostic marker \cite{ryan_breast_2017, yurdaisik_evaluation_2020}. These findings have created interest in development of automated computer vision tools to detect BAC using already acquired digital mammograms for cancer screening \cite{guo_scu-net_2021, urooj_multi-task_2024}.
\subsection{Existing AI-based models for BAC}
Our initial research established a robust patch-based segmentation model for BAC detection that effectively handles the high dimensionality of mammograms \cite{guo_scu-net_2021}. This model included several quantification methods, such as measuring the total area of calcified vessels and calculating the total area of calcified pixels above specific intensity thresholds. We also recently demonstrated that BAC is prognostic for CVD in a single center study \cite{urooj_multi-task_2023}. BAC scores were dichotomized into low BAC, mid BAC, and high BAC where the survival time is considered between the screening date and the adverse cardiac event. We found the survival probability of low BAC subjects to be $\sim$93\% even after 15 years, whereas the probability of not developing CVD for high BAC subjects dropped to 68\%-75\% in 15 years. A recent single-center study using a commercially available tool demonstrated that BAC grading enhanced CVD prediction beyond standard risk factors \cite{allen_automated_2024}. However, this study had a paucity of African American patients who are known to exhibit different rates of CVD \cite{tsao_heart_2022}, imputed missing data, and did not mandate a follow-up period. 
Our study assesses whether automatically quantified BAC from routine screening mammograms predicts cardiovascular outcomes in a large, multi-institutional, racially diverse population. This automated approach enables opportunistic cardiovascular risk assessment in women already undergoing mammography screening, potentially identifying those at increased risk without additional testing.

\section{Methods}
\subsection{Ethical Approvals}
This study complied with all relevant ethical regulations. Study protocols received institutional review board approval at each study site, and a waiver of consent was granted due to the retrospective nature of this study. This study followed the Strengthening the Reporting of Observational Studies in Epidemiology (STROBE) reporting guidelines.

\subsection{Study Population}
\begin{figure}[htb]
\centering
\includegraphics[width=0.95\linewidth]{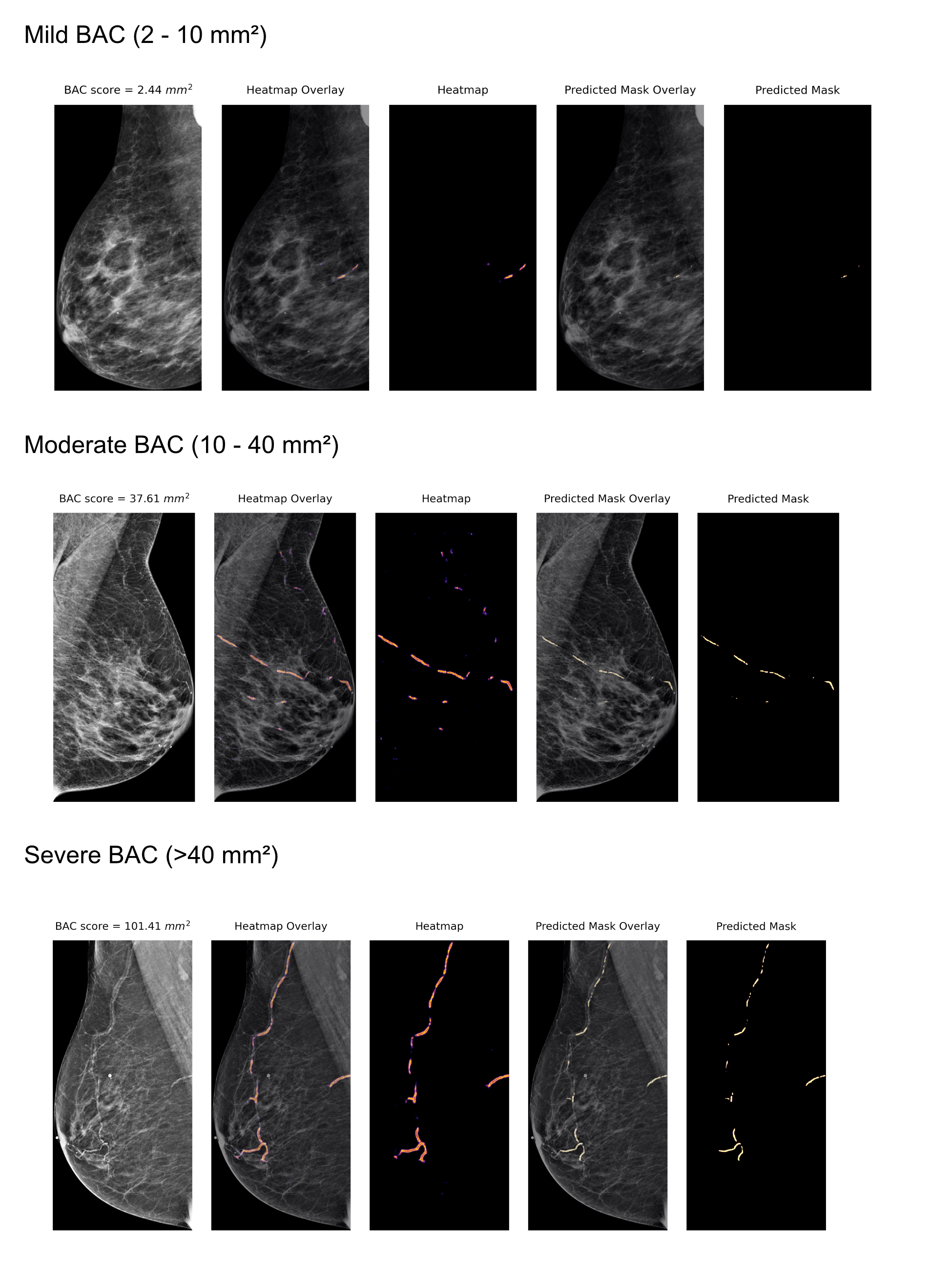}
\caption{Examples of mammograms with mild, moderate and severe BAC quantified by our AI model.}
\label{fig:bac_examples}
\end{figure}

The study population is summarized in Figure \ref{fig:bac_examples}. Our study included 54,743 women from Emory Healthcare (Atlanta, GA, USA) in the Emory Breast Imaging Dataset (EMBED) \cite{jeong_emory_2023} who served as our internal cohort for development and validation, and 61,392 women from Mayo Clinic Enterprise (Arizona, Florida, Mid-West, Rochester) who constituted our external validation cohort (Supplemental Figure \ref{fig:patient_flow}). Inclusion criteria included age $>18$ years, at least one CVD event following an initial screening mammogram, or at least 5 years of face-to-face clinical follow-up in women who had no events. For each patient from Emory Healthcare, the earliest screening mammogram was included for evaluation to maximize the capture window for events and follow-up period. At Mayo Clinic, all mammogram samples were from 2017. Therefore, each patient was represented only as a single sample at both sites. 
To maximize exclusion of women at Emory who had known cardiovascular disease at the time of screening, we also excluded patients who had existing records of Major Adverse Cardiovascular Events (MACE) or existing CTs for coronary calcium or coronary angiography before the index mammogram.

\subsection{Measurement of Breast Arterial Calcification}
\subsubsection{Model Development}
To increase the BAC segmentation accuracy over the previously published patch-based segmentation model \cite{guo_scu-net_2021} which does not fully leverage the spatial location of the patches, we developed a proprietary transformer-based neural network architecture to directly segment and quantify BAC within the mediolateral oblique (MLO) view of high resolution 2D full-field digital screening mammograms (FFDM). The proposed model is a semantic segmentation model which includes a hierarchical transformer encoder to generate low and high dimensional embedding from high resolution FFDM and light weighted decoder to fuse multi-level features. The model was trained on expert-annotated screening mammograms by a trained human annotator (HT) and divided into 90\%-5\%-5\% training, validation, and test exams.
The final model output is a binary mask with detected BAC pixels. BAC area is quantified by computing the number of segmented pixels multiplied by the pixel size defined in the DICOM metadata to yield an area in mm$^2$. Model performance compared to expert-annotation at the pixel-level had 68.3\% accuracy, 0.76 r$^2$ score and 61 mean Intersection over Union (mIoU) score, which supersedes previously reported inter-observer variability for BAC which ranges from 0.48-0.53 \cite{hendriks_breast_2015}.
\subsubsection{BAC Severity Classification}
BAC severity was categorized into four groups based on the quantified BAC area: No BAC (0-2 mm$^2$), Mild (2-10 mm$^2$), Moderate (11-40 mm$^2$) and Severe (>40 mm$^2$) (Figure 1). These thresholds were established through empirical analysis that maximized between-group hazard ratio differences in an independent Emory patient cohort (Supplemental Figure \ref{fig:bac_empirical_severity_classification}). The No BAC category’s upper threshold of 2 mm$^2$ was chosen to account for potential image noise artifacts, which was validated through visual inspection of 100 exams.

\subsection{Identification of Cardiovascular Events, Risk Factors, and ASCVD Score}
Expert-curated ICD codes were extracted from the electronic health record (EHR) with timestamp data from both clinical sites to identify MACE \cite{bosco_major_2021} (Supplemental Table \ref{tab:icd}) and all-cause mortality. For women without events at Emory, a minimum five-year follow-up from mammogram date was required, defined by an in-person visit to a healthcare provider at the institution. Mayo Clinic women had minimum 6-year follow-up.
To compare BAC to the ASCVD risk score \cite{goff_2013_2014}, risk factors including diabetes, hypertension, use of hypertensive medication, lipid levels, A1C, age, and blood pressure were extracted from the EHR as ICD codes or lab values. Only 5,180 of the 54,753 Emory patients (9,5\%) had all laboratory values available to calculate the ASCVD score.
Emory patients’ ASCVD scores were calculated by implementing the algorithm from Goff, et al \cite{goff_2013_2014}. We validated our implementation by comparing computed scores for 40 random patients against the online American College of Cardiology ASCVD Risk Estimator Plus calculator \cite{american_college_of_cardiology_ascvd_nodate} using manually entered data (R=0.99, Supplemental Figure \ref{fig:ascvd_validate}). The ASCVD score calculator was then applied to all eligible Emory patients with complete risk factor data. For Mayo patients, ASCVD score was obtained from the cardiology registry which was expert-curated for clinical assessment.

\subsection{Statistical Analysis}
We compared the prevalence of various demographics and risk factors between BAC severity groups and presence or absence of MACE using several statistical tests: chi-square for discrete variables and t-tests or one-way ANOVA for continuous variables. The Kruskal-Wallis test was used to assess the differences in medians follow-up duration between groups.
Kaplan-Meier curves were generated to analyze the association between BAC severity and MACE occurrence following a patient’s mammograms. Primary analyses were conducted in two main age groups: 40-60 years and 60-80 years. Given the potential prognostic value in younger populations, we also performed a dedicated analysis in patients under 50 years of age, specifically examining MACE outcomes in those with any detectable BAC compared to those without BAC, regardless of severity scoring. Differences in event-free survival rates between BAC severity groups were assessed using the multivariate log-rank test.
Cox proportional-hazards models were used to assess how BAC predicted MACE events. We examined BAC separately as categorical and continuous variables. For the continuous analysis, we transformed BAC measurements using $\log_2$(BAC+1). This transformation, similar to that used in coronary artery calcium studies \cite{detrano_coronary_2008, mcclelland_10-year_2015}, allows us to examine how doubling the BAC measurement affects event risk. Adding 1 to BAC before transformation enables us to include zero BAC values in our analysis.
We adjusted the Cox proportional hazards models for age, diabetes, smoking, and medication use (statins and antihypertensives).  Subsequently, we developed a separate model that included the ASCVD risk score as a covariate, allowing us to account for traditional cardiovascular risk factors and assess the independent effect of BAC.

\section{Results}
The characteristics of the study population are shown in Table \ref{tab:demographics}. The presence of diabetes, use of antihypertensive medications and statins, smoking, systolic blood pressure, and body mass index and lower eGFR all correlated positively with the severity of BAC and MACE.

\begin{table*}[htb]
\centering
\caption{Baseline demographic characteristics and risk factors based on the BAC severity and whether a study participant had any subsequent MACE event}
\label{tab:demographics}
\resizebox{0.95\textwidth}{!}{%
\begin{tabular}{lccccccccc}
\hline
\multirow{2}{*}{} & \multicolumn{5}{c}{BAC Severity} & \multicolumn{3}{c}{MACE} & {All Participants} \\
\cmidrule(lr){2-6} \cmidrule(lr){7-9}
 & Zero & Mild & Moderate & Severe & {p} & No Event & Event & {p} & \\
 & (N=36,333) & (N=13,345) & (N=3,914) & (N=1,161) & & (N=50,321) & (N=4,432) & & (N=54,753) \\
 & [66.4\%] & [24.4\%] & [7.1\%] & [2.1\%] & & [92.0\%] & [8.0\%] & & \\
\hline
\multicolumn{10}{c}{\textbf{Internal Dataset}} \\
\hline
Age, y (std) & 55.6 (10.0) & 58.5 (10.3) & 61.5 (10.5) & 68.3 (9.0) & <0.001 & 56.4 (10.2) & 63.8 (9.9) & <0.001 & 57.0 (10.4) \\
Race (\%) & & & & & 0.99 & & & 0.44 & \\
\quad Asian & 6.6 & 2.9 & 2.7 & 3.8 & & 5.6 & 2.9 & & 5.4 \\
\quad Black & 43.7 & 54.1 & 54.5 & 55.5 & & 46.2 & 59.4 & & 47.2 \\
\quad Other & 6.6 & 5.3 & 4.4 & 4.5 & & 6.4 & 2.6 & & 6.1 \\
\quad White & 43.1 & 37.7 & 38.3 & 36.2 & & 41.8 & 35.2 & & 41.3 \\
Hispanic (\%) & 4.4 & 3.1 & 3.1 & 4.3 & 0.89 & 4.1 & 2.4 & 0.12 & 4.0 \\
Median time to follow up, y & 8.0 & 8.0 & 7.5 & 7.0 & <0.001 & 8.0 & 4.0 & <0.001 & 8.0 \\
Diabetes (\%)$^\dagger$ & 10.25 & 20.0 & 27.0 & 41.0 & <0.001 & 12.5 & 36.0 & <0.001 & 14.5 \\
Antihypertensives (\%)$^\dagger$ & 4.4 & 7.7 & 11.2 & 21.8 & <0.001 & 4.4 & 25.0 & <0.001 & 5.0 \\
Statins (\%)$^\dagger$ & 3.5 & 6.4 & 10.2 & 20.2 & <0.001 & 3.6 & 20.7 & <0.001 & 4.9 \\
Smoking (\%)$^\dagger$ & 7.6 & 10.9 & 12.7 & 15.8 & <0.001 & 8.0 & 19.9 & <0.001 & 8.9 \\
Total Cholesterol, mg/dL (std)$^\dagger$ & 191.8 (32.9) & 190.7 (34.5) & 187.8 (37.1) & 190.0 (37.2) & 0.14 & 191.4 (33.5) & 189.4 (35.9) & 0.16 & 191.2 (33.8) \\
HDL, mg/dL (std)$^\dagger$ & 60.4 (15.2) & 57.3 (14.2) & 57.2 (14.0) & 59.3 (14.6) & <0.001 & 59.6 (14.9) & 57.6 (14.7) & 0.002 & 59.3 (14.9) \\
Systolic Blood Pressure (mmHg) (std)$^\dagger$ & 127.2 (20.5) & 131.7 (20.1) & 134.3 (20.5) & 137.4 (22.9) & <0.001 & 128.4 (20.0) & 136.6 (24.0) & <0.001 & 129.0 (20.6) \\
BMI, kg/mm$^2$ (std)$^\dagger$ & 28.3 (6.8) & 32.4 (7.8) & 34.2 (9.3) & 31.5 (7.9) & <0.001 & 29.7 (7.5) & 31.3 (8.2) & <0.001 & 28.5 (7.6) \\
eGFR, mL/min/1.73m$^2$ (std)$^\dagger$ & 68.9 (20.0) & 64.9 (20.1) & 61.6 (20.8) & 48.1 (23.8) & <0.001 & 68.4 (19.6) & 54.6 (22.6) & <0.001 & 66.4 (21.1) \\
\hline
\multicolumn{10}{c}{\textbf{External Dataset}} \\
\hline
 & (N=39,920) & (N=15,466) & (N=7,523) & (N=3,334) & {p} & (N=59,204) & (N=7,023) & {p} & (N=61,390) \\
 & [65.13\%] & [25.23\%] & [12.27\%] & [5.43\%] & & [89.40\%] & [10.60\%] & & \\
Age, y (std) & 52.99 (21.09) & 53.95 (20.56) & 55.36 (19.85) & 57.86 (18.38) & & & & & 59.78 (11.64) \\
Race (\%) & & & & & & & & & \\
\quad Asian & 2.19 & 1.33 & 0.85 & 1.41 & & 1.90 & 0.98 & & 1.80 \\
\quad Black & 0.93 & 1.24 & 1.13 & 1.47 & & 1.06 & 0.98 & & 1.05 \\
\quad Other & 1.97 & 1.88 & 1.58 & 1.68 & & 2.01 & 0.89 & & 1.89 \\
\quad White & 94.89 & 95.5 & 96.43 & 95.42 & & 95.02 & 97.13 & & 95.24 \\
Hispanic (\%) & 1.92 & 1.64 & 1.26 & 1.68 & & 1.84 & 1.13 & & 1.77 \\
Diabetes (\%) & 17.08 & 42.23 & 72.15 & 68.98 & <0.001 & 10.86 & 34.41 & <0.01 & 14.44 \\
Antihypertensives (\%)$^\dagger$ & 0.7 & 1.58 & 2.45 & 2.49 & <0.001 & 0.32 & 2.19 & <0.001 & 0.52 \\
Statins (\%)$^\dagger$ & 3.36 & 8.94 & 14.59 & 16.73 & <0.001 & 1.44 & 14.47 & <0.001 & 2.82 \\
Total Cholesterol, mg/dL (std)$^\dagger$ & 198.58 (43.11) & 195.57 (44.2) & 192.63 (43.74) & 184.79 (43.63) & & 198 (42.99) & 183.17 (44.96) & & 195.8 (43.7) \\
HDL, mg/dL (std)$^\dagger$ & 65.96 (20.44) & 62.98 (19.29) & 60.98 (18.22) & 60.49 (18.62) & & 64.68 (19.72) & 60.02 (19.56) & & 63.9 (19.77) \\
Systolic Blood Pressure (mmHg) (std)$^\dagger$ & 130.2 (19.5) & 133.2 (19.1) & 134.5 (19.3) & 136.4 (21.9) & & 131.4 (19.2) & 136.4 (21.5) & & 131.3 (19.5) \\
BMI, kg/mm$^2$ (std)$^\dagger$ & 23.4 (10.4) & 25.5 (9.8) & 28.2 (10.5) & 27.4 (10.8) & & 24.2 (10.2) & 27.4 (10.8) & & 24.7 (9.8) \\
eGFR, mL/min/1.73m$^2$ (std)$^\dagger$ & 103.4 (19.3) & 98.7 (19.2) & 93.3 (19.5) & 84.5 (19.1) & <0.001 & 104.5 (19.5) & 86.5 (20.1) & <0.001 & 93.2 (20.4) \\
\hline
\multicolumn{10}{l}{$^\dagger$ Calculated on patients who have the corresponding value} \\
\end{tabular}}
\end{table*}

BAC severity increased with age, with approximately 13 years age difference between patients with No BAC compared to those with Severe BAC (p<0.001). The mean age of patients who experienced MACE was also 8 years higher than those who did not (p<0.001). Overall, 66.4\%, 24.4\%, 7.1\%, and 2.1\% of patients had No BAC, Mild, Moderate, and Severe BAC, respectively.
\subsection{BAC as Risk Factor for MACE}
\begin{figure*}[h!tb]
\centering
\includegraphics[width=0.95\linewidth]{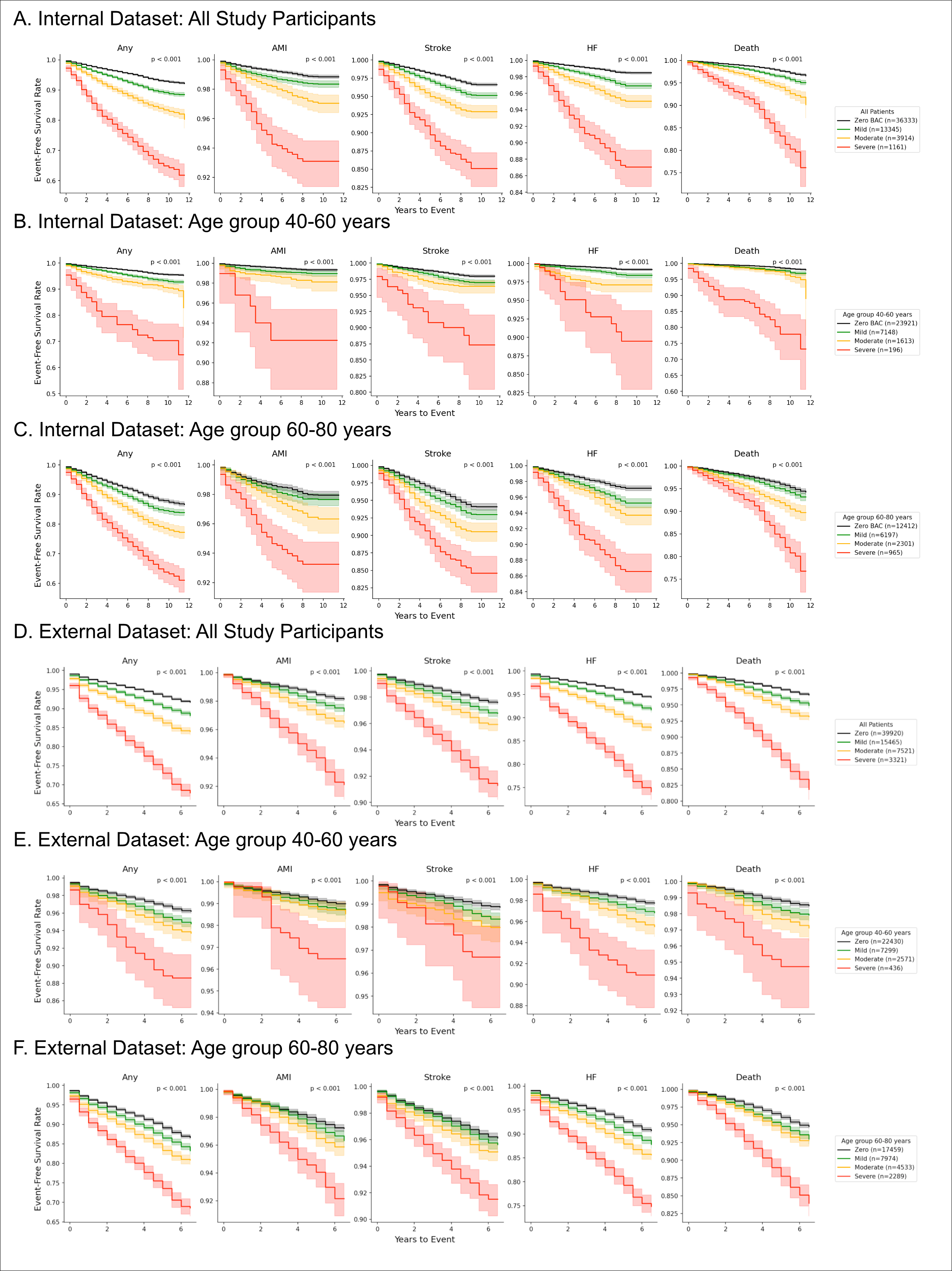}
\caption{A, D: Kaplan-Meier curves of study participants across any MACE event and each MACE event by BAC severity; Further stratification by age groups 40-60 years (B, E) and 60-80 years (C, F). A, B, C are curves for the internal dataset while D, E and F are curves for the external dataset.}
\label{fig:km_all_age_groups}
\end{figure*}

Any amount of BAC was associated with increased risk of MACE across internal and external datasets. In the internal dataset, after adjusting for CVD risk factors including age, diabetes, smoking, statin use, and antihypertensive medication use, Mild (HR 1.18, 95\% CI 1.10-1.27, p<0.001), Moderate (HR 1.47, 95\% CI 1.33-1.61, p<0.001), and Severe BAC (HR 2.22, 95\% CI 1.98-2.50, p<0.001) were independently associated with MACE. Similar findings were observed in the external dataset (Mild: HR 1.22, 95\% CI 1.15-1.30, p<0.001; Moderate: HR 1.38, 95\% CI 1.29-1.48, p<0.001; Severe: HR 2.03, 95\% CI 1.88-2.19, p<0.001) (Table \ref{tab:bac_clinical_outcomes}). When treating BAC as a continuous variable, these findings were further supported, showing consistent associations with MACE in both internal (HR 1.13, 95\% CI 1.11-1.15, p<0.001) and external (HR 1.15, 95\% CI 1.13-1.16, p<0.001) datasets.
Kaplan-Meier curves in both datasets demonstrated clear separation between No BAC, Mild, Moderate, and Severe BAC for all outcomes (AMI, stroke, heart failure, and all-cause mortality), with non-overlapping confidence intervals (Figure \ref{fig:km_all_age_groups}). This relationship persisted across age groups, with similar patterns observed in both younger (age 40-60) and older (age 60-80) patients. Notably, in patients under 50 years old (Figure \ref{fig:km_50_ascvd}A), where BAC is less frequently observed, the presence of any BAC was associated with significantly lower event-free survival for MACE in both internal and external cohorts (p<0.001 for both), underscoring that even slight BAC in this younger population may indicate increased cardiovascular risk.

\begin{table}[htbp]
\centering
\caption{Association of BAC and clinical outcomes}
\label{tab:bac_clinical_outcomes}
\resizebox{0.95\textwidth}{!}{%
\begin{tabular}{lccccc}
\hline
\multirow{2}{*}{} & \textbf{Composite MACE} & \textbf{AMI} & \textbf{Stroke} & \textbf{HF} & \textbf{Death} \\
\cmidrule(lr){2-6}
 & HR (95\%CI) & HR (95\%CI) & HR (95\%CI) & HR (95\%CI) & HR (95\%CI) \\
 & p-value & p-value & p-value & p-value & p-value \\
\hline
\multicolumn{6}{l}{\textbf{Internal}} \\
\hline
\multicolumn{6}{l}{\textit{Unadjusted Hazard Ratios}} \\
\textbf{Zero BAC} & Reference & Reference & Reference & Reference & Reference \\
\textbf{Mild BAC} & 1.59 (1.48-1.70) & 1.50 (1.26-1.78) & 1.50 (1.35-1.66) & 2.00 (1.74-2.30) & 1.60 (1.42-1.80) \\
 & <0.001 & <0.001 & <0.001 & <0.001 & <0.001 \\
\textbf{Moderate BAC} & 2.57 (2.34-2.81) & 2.54 (2.03-3.18) & 2.31 (2.01-2.66) & 3.37 (2.82-4.02) & 2.77 (2.38-3.22) \\
 & <0.001 & <0.001 & <0.001 & <0.001 & <0.001 \\
\textbf{Severe BAC} & 6.11 (5.47-6.82) & 6.69 (5.20-8.60) & 5.03 (4.23-5.97) & 9.41 (7.73-11.45) & 7.78 (6.60-9.18) \\
 & <0.001 & <0.001 & <0.001 & <0.001 & <0.001 \\
\textbf{log2(BAC+1)} & 1.33 (1.30-1.35) & 1.33 (1.28-1.38) & 1.29 (1.26-1.32) & 1.44 (1.40-1.48) & 1.37 (1.34-1.40) \\
 & <0.001 & <0.001 & <0.001 & <0.001 & <0.001 \\
\hline
\multicolumn{6}{l}{\textit{Adjusted Hazard Ratios*}} \\
\textbf{Zero BAC} & Reference & Reference & Reference & Reference & Reference \\
\textbf{Mild BAC} & 1.18 (1.10-1.27) & 1.08 (0.91-1.29) & 1.13 (1.02-1.26) & 1.37 (1.19-1.57) & 1.19 (1.06-1.34) \\
 & <0.001 & 0.37 & 0.02 & <0.001 & <0.005 \\
\textbf{Moderate BAC} & 1.47 (1.33-1.61) & 1.38 (1.10-1.73) & 1.35 (1.17-1.56) & 1.67 (1.39-2.00) & 1.61 (1.38-1.88) \\
 & <0.001 & 0.01 & <0.001 & <0.001 & <0.001 \\
\textbf{Severe BAC} & 2.22 (1.98-2.50) & 2.18 (1.67-2.85) & 1.85 (1.55-2.22) & 2.66 (2.16-3.27) & 2.83 (2.38-3.37) \\
 & <0.001 & <0.001 & <0.001 & <0.001 & <0.001 \\
\textbf{log2(BAC+1)} & 1.13 (1.11-1.15) & 1.13 (1.08-1.17) & 1.10 (1.07-1.13) & 1.17 (1.14-1.21) & 1.17 (1.14-1.20) \\
 & <0.001 & <0.001 & <0.001 & <0.001 & <0.001 \\
\hline
\multicolumn{6}{l}{\textbf{External}} \\
\hline
\multicolumn{6}{l}{\textit{Unadjusted Hazard Ratios}} \\
\textbf{Zero BAC} & Reference & Reference & Reference & Reference & Reference \\
\textbf{Mild BAC} & 1.42 (1.34-1.51) & 1.39 (1.23-1.58) & 1.35 (1.21-1.51) & 1.49 (1.38-1.59) & 1.48 (1.35-1.62) \\
 & <0.001 & <0.001 & <0.001 & <0.001 & <0.001 \\
\textbf{Moderate BAC} & 2.06 (1.93-2.21) & 1.90 (1.65-2.20) & 1.75 (1.54-2.00) & 2.30 (2.13-2.49) & 2.11 (1.91-2.34) \\
 & <0.001 & <0.001 & <0.001 & <0.001 & <0.001 \\
\textbf{Severe BAC} &  &  &  &  &  \\
 & <0.001 & <0.001 & <0.001 & <0.001 & <0.001 \\
\textbf{log2(BAC+1)} & 1.32 (1.31-1.34) & 1.29 (1.26-1.32) & 1.26 (1.24-1.29) & 1.37 (1.35-1.39) & 1.36 (1.34-1.39) \\
 & <0.001 & <0.001 & <0.001 & <0.001 & <0.001 \\
\hline
\multicolumn{6}{l}{\textit{Adjusted Hazard Ratios*}} \\
\textbf{Zero BAC} & Reference & Reference & Reference & Reference & Reference \\
\textbf{Mild BAC} & 1.22 (1.15-1.30) & 1.22 (1.08-1.38) & 1.17 (1.05-1.31) & 1.26 (1.17-1.35) & 1.25 (1.14-1.37) \\
 & <0.001 & <0.005 & 0.01 & <0.001 & <0.001 \\
\textbf{Moderate BAC} & 1.38 (1.29-1.48) & 1.34 (1.16-1.56) & 1.21 (1.06-1.39) & 1.49 (1.38-1.62) & 1.37 (1.24-1.53) \\
 & <0.001 & <0.001 & <0.005 & <0.001 & <0.001 \\
\textbf{Severe BAC} & 2.03 (1.88-2.19) & 2.17 (1.85-2.54) & 1.81 (1.57-2.09) & 2.21 (2.02-2.41) & 2.40 (2.15-2.67) \\
 & <0.001 & <0.001 & <0.001 & <0.001 & <0.001 \\
\textbf{log2(BAC+1)} & 1.15 (1.13-1.16) & 1.14 (1.11-1.17) & 1.11 (1.09-1.14) & 1.17 (1.16-1.19) & 1.17 (1.15-1.19) \\
 & <0.001 & <0.001 & <0.001 & <0.001 & <0.001 \\
\hline
\multicolumn{6}{l}{* Adjusted for age, sex, race, systolic blood pressure, diabetes, smoking, total cholesterol, HDL, and eGFR.} \\
\end{tabular}}
\end{table}

\subsection{BAC compared to ASCVD Scores}
\begin{figure*}[h!tb]
\centering
\includegraphics[width=0.95\linewidth]{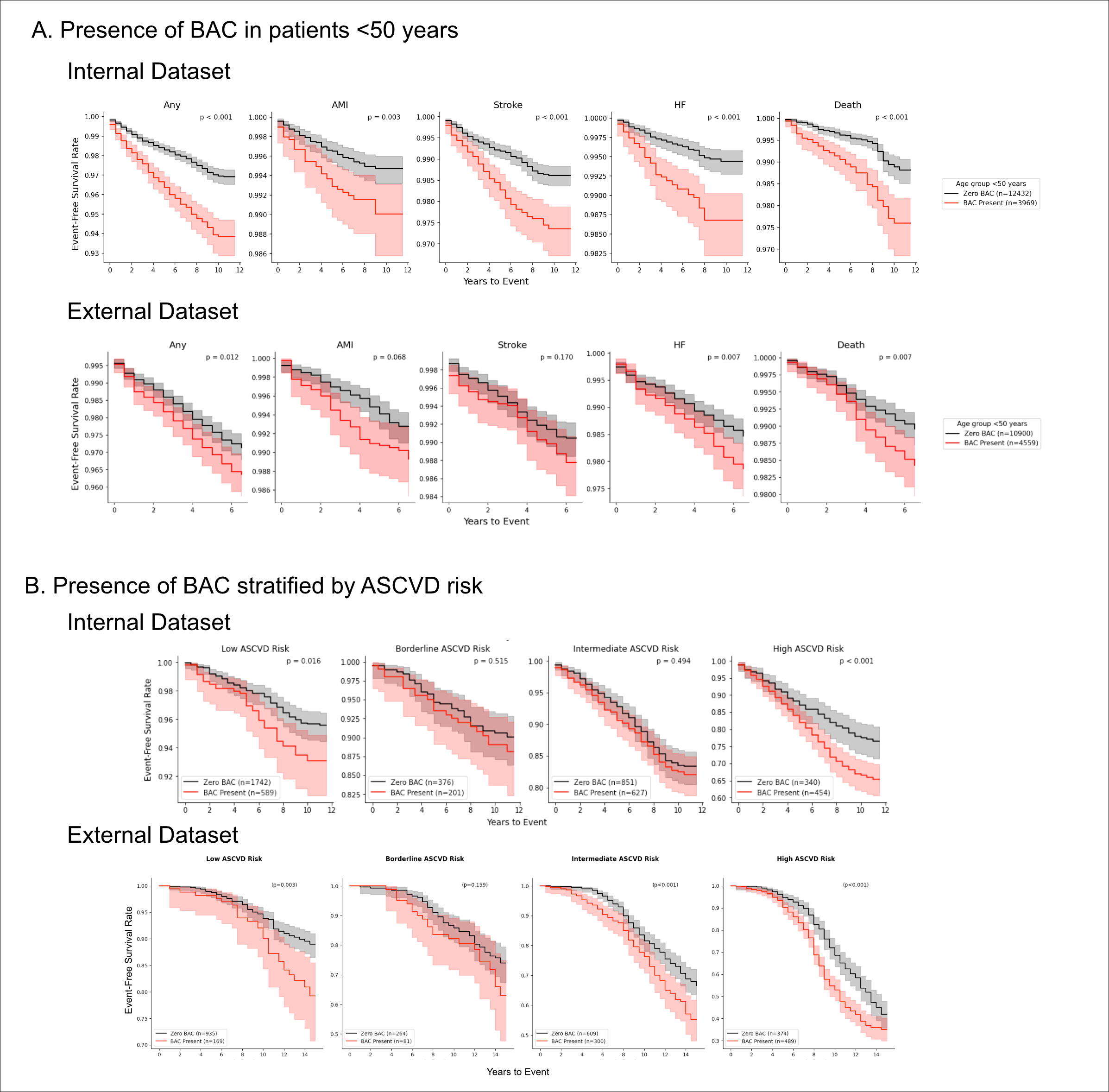}
\caption{The relationship between MACE events with the presence or absence of BAC. A. In patients <50years; B. In patient groups stratified by their ASCVD risk category.}
\label{fig:km_50_ascvd}
\end{figure*}

ASCVD risk score was calculated in 5,180 patients for whom all risk data was available in the internal dataset. In the external validation dataset, extant ASCVD risk score was available for 3,221 patients. Kaplan-Meier survival analysis stratified by ASCVD risk categories (Figure \ref{fig:km_50_ascvd}B) demonstrated that the presence of BAC was associated with significantly lower event-free survival for any MACE events in both Low-Risk (p=0.016) and High-Risk (p<0.001) patients, but not in Borderline and Intermediate risk patients. In the external validation cohort, BAC’s prognostic significance was demonstrated for Low  (p=0.003), Intermediate (p<0.001), and High-Risk (p<0.001) patients.
Multivariate Cox regression analysis treating ASCVD score and BAC as continuous variables revealed that BAC was still an independent predictor of MACE outcomes. In the internal cohort, BAC demonstrated significant associations with each MACE outcome: AMI (HR 1.18, 95\% CI 1.06-1.31, p<0.005), stroke (HR 1.10, 95\% CI 1.03-1.18, p<0.005), HF (HR 1.27, 95\% CI 1.18-1.37, p<0.001), and all-cause mortality (HR 1.17, 95\% CI 1.09-1.25, p<0.001). BAC also showed significant predictive value for the composite any MACE outcome (HR 1.13, 95\% CI 1.09-1.19), p<0.001). Interestingly, significantly stronger hazard ratios were observed in the external validation cohort: AMI (HR 1.22, 95\% CI 1.16-1.28, p<0.001), stroke (HR 1.23, 95\% CI 1.17-1.29, p<0.001), HF (HR 1.22, 95\% CI 1.19-1.26, p<0.001), all-cause mortality (HR 1.30, 95\% CI 1.26-1.34, p<0.001), and any MACE event (HR 1.27, 95\% CI 1.23-1.30), p<0.001).

\section{Discussion}
Our results demonstrate that BAC score is a strong and independent predictor of adverse cardiovascular events, even when stratified by patient age, known cardiovascular risk factors, and ASCVD scores. This is in line with existing literature that has demonstrated the prognostic value of BAC for CVD \cite{allen_automated_2024, hendriks_breast_2015, rotter_breast_2008, schnatz_association_2011, iribarren_breast_2022}, perhaps due to the underlying pathophysiology of BAC which is different than that of atherosclerosis \cite{duhn_breast_2011, manzoor_progression_2018, oneill_breast_2014}.
The prevalence of BAC was similar across both institutions, ranging from 33.3\% to 34.9\%. These values are slightly higher than the 29.4\% prevalence reported by Lanzer, et al \cite{lanzer_medial_2014}. Other studies have found significantly lower prevalence rates, including 14\% \cite{rotter_breast_2008} and 20.8\% \cite{schnatz_association_2011}. This difference may be attributed to our model's  higher sensitivity towards BAC detection, where we classify any calcification greater than 2 mm$^2$ as BAC positive.
Our study represents a significant advancement in BAC assessment by implementing an objective, automated quantification system, moving beyond the traditional subjective radiologist scoring that has shown only moderate inter-observer agreement ($\kappa$ = 0.48-0.53) \cite{hendriks_breast_2015}. A recent large-scale single-center study by Allen, et al \cite{allen_automated_2024}, utilized an AI-based relative scoring system (0-100) to calculate BAC severity, however their study was single-center, contained few African American patients, and considered BAC score by age-based quartiles rather than absolute score. Instead, we quantifying the absolute BAC measurements (<2 mm$^2$, 2-10 mm$^2$, 10-40 mm$^2$, and >40 mm$^2$, respectively) which offers potential advantages in clinical interpretation and standardization across institutions and populations, as it provides absolute measurements of BAC area rather than relative scores.
This quantitative approach enables analysis of relationships between clinical risk factors and BAC severity. In our internal dataset, we observed that as BAC severity increased, there were significantly higher proportions of patients with diabetes mellitus, smoking history, and use of antihypertensive medications and statins. Systolic blood pressure also showed a positive correlation with BAC severity. These findings align with previous studies that reported higher prevalence of these risk factors in BAC-positive patients \cite{rotter_breast_2008, bae_association_2013}. We also observed that BAC severity correlated with higher smoking rates, contrasting with Hendriks et al. who reported that smoking was associated with lower BAC prevalence \cite{hendriks_breast_2015}. Their study also found no associations between BAC and hypertension, obesity, or dyslipidemia. This difference may be attributed to population heterogeneity or explained by the treatment attenuating risk.
Furthermore, BAC remains a statistically significant prognosticator for MACE when included in multivariate adjustment with ASCVD scores, suggesting that it can be used to further risk stratify patients. In Emory population of 54,743 women, data required calculate ASCVD scores was only available in the EHR for 5,180 women (9.5\%), even when including a study population that had a minimum of 5 years of clinical follow-up at Emory. This is corroborated by existing literature that shows that the data required for CVD risk prediction tools is still not available in the majority of the population domestically and internationally \cite{borden_reducing_2022}. In contrast, screening mammography is performed on 40 million women annually in the United States \cite{american_cancer_society_cancer_2024} and is a strong predictor of CVD without the need for additional clinical data. This can allow capture of high-risk patients during routine mammography visits and referral to cardiology, which generally show strong adherence.
Our analysis revealed that breast arterial calcification (BAC) has significant prognostic value in multiple risk groups. In both datasets, BAC showed strong predictive power for low-risk (p=0.016) and high-risk (p<0.001) populations. For intermediate-risk patients, BAC's prognostic value was significant (p<0.001) in the external dataset. These findings point to two important clinical applications. First, BAC can identify patients at high CVD risk who would be classified as low risk by traditional scoring. This is particularly relevant since the ASCVD risk score is known to underestimate cardiovascular risk in minority groups and patients of lower socioeconomic status \cite{lloyd-jones_use_2019}. BAC identified elevated risk in 33\% of patients in our internal dataset and 18\% in our external dataset who were classified as low-risk by ASCVD risk score. Second, BAC provides additional risk stratification for intermediate and high-risk patients, helping identify candidates for intensive cardiac evaluation. Future studies should evaluate BAC with the PREVENT risk score \cite{khan_development_2024}, which incorporates socioeconomic factors, potentially providing a more comprehensive and equitable foundation for leveraging opportunistic BAC findings in cardiovascular risk assessment.
Finally, we demonstrate that our deep learning-based tool for automated BAC quantification is robust across a demographically diverse, multi-site screening mammography population. The model performed well across Hologic and GE scanners and is robust despite many mammographic artifacts such as mole markers, biopsy clips, scar markers, and breast implants (Supplemental Figure \ref{fig:artifacts}). Occasionally, we observed false positives for very dense ductal calcifications (Supplemental Figure \ref{fig:false_positives}), however this was infrequent and can easily be distinguished from BAC.

\subsection{Limitations}
This study is conducted using retrospective data from two US institutions and two scanner manufacturers, and further work is needed to ensure performance of the model across all scanners. MACE was extracted from the EHR using previously validated diagnostic and procedural codes \cite{bosco_major_2021}, but manual chart review or contact of patients for verification of outcomes was not practical given the large cohort and relatively low prevalence of disease. Cardiovascular mortality is unreliable from EHR data, so all-cause mortality was used instead. Finally, the model functions on full-field digital mammography (FFDM) which is a traditional 2D method of breast imaging. Digital breast tomosynthesis (DBT) is being increasingly used in the United States instead of FFDM \cite{richman_adoption_2019, lee_effect_2018}, and therefore a BAC quantification model for DBT is needed to maximize capture of high-risk patients at mammography.

\section{Conclusion}
Automated BAC quantification from routine mammography provides an opportunistic and effective cardiovascular risk assessment method in women, without additional radiation exposure. Its predictive value is independent of traditional risk factors and ASCVD scores, particularly in younger women, suggesting that BAC screening could be valuable for early cardiovascular risk detection in women undergoing routine mammography.

\printbibliography

@article{goff_2013_2014,
	title = {2013 {ACC}/{AHA} {Guideline} on the {Assessment} of {Cardiovascular} {Risk}: {A} {Report} of the {American} {College} of {Cardiology}/{American} {Heart} {Association} {Task} {Force} on {Practice} {Guidelines}},
	volume = {129},
	issn = {0009-7322, 1524-4539},
	shorttitle = {2013 {ACC}/{AHA} {Guideline} on the {Assessment} of {Cardiovascular} {Risk}},
	url = {https://www.ahajournals.org/doi/10.1161/01.cir.0000437741.48606.98},
	doi = {10.1161/01.cir.0000437741.48606.98},
	language = {en},
	number = {25\_suppl\_2},
	urldate = {2024-10-31},
	journal = {Circulation},
	author = {Goff, David C. and Lloyd-Jones, Donald M. and Bennett, Glen and Coady, Sean and D’Agostino, Ralph B. and Gibbons, Raymond and Greenland, Philip and Lackland, Daniel T. and Levy, Daniel and O’Donnell, Christopher J. and Robinson, Jennifer G. and Schwartz, J. Sanford and Shero, Susan T. and Smith, Sidney C. and Sorlie, Paul and Stone, Neil J. and Wilson, Peter W. F.},
	month = jun,
	year = {2014},
}

@article{detrano_coronary_2008,
	title = {Coronary {Calcium} as a {Predictor} of {Coronary} {Events} in {Four} {Racial} or {Ethnic} {Groups}},
	volume = {358},
	issn = {0028-4793},
	url = {https://www.nejm.org/doi/full/10.1056/NEJMoa072100},
	doi = {10.1056/NEJMoa072100},
	number = {13},
	urldate = {2024-10-11},
	journal = {New England Journal of Medicine},
	author = {Detrano, Robert and Guerci, Alan D. and Carr, J. Jeffrey and Bild, Diane E. and Burke, Gregory and Folsom, Aaron R. and Liu, Kiang and Shea, Steven and Szklo, Moyses and Bluemke, David A. and O'Leary, Daniel H. and Tracy, Russell and Watson, Karol and Wong, Nathan D. and Kronmal, Richard A.},
	month = mar,
	year = {2008},
	pages = {1336--1345},
}

@article{mcclelland_10-year_2015,
	title = {10-{Year} {Coronary} {Heart} {Disease} {Risk} {Prediction} {Using} {Coronary} {Artery} {Calcium} and {Traditional} {Risk} {Factors}: {Derivation} in the {MESA} ({Multi}-{Ethnic} {Study} of {Atherosclerosis}) {With} {Validation} in the {HNR} ({Heinz} {Nixdorf} {Recall}) {Study} and the {DHS} ({Dallas} {Heart} {Study})},
	volume = {66},
	issn = {1558-3597},
	shorttitle = {10-{Year} {Coronary} {Heart} {Disease} {Risk} {Prediction} {Using} {Coronary} {Artery} {Calcium} and {Traditional} {Risk} {Factors}},
	doi = {10.1016/j.jacc.2015.08.035},
	language = {eng},
	number = {15},
	journal = {Journal of the American College of Cardiology},
	author = {McClelland, Robyn L. and Jorgensen, Neal W. and Budoff, Matthew and Blaha, Michael J. and Post, Wendy S. and Kronmal, Richard A. and Bild, Diane E. and Shea, Steven and Liu, Kiang and Watson, Karol E. and Folsom, Aaron R. and Khera, Amit and Ayers, Colby and Mahabadi, Amir-Abbas and Lehmann, Nils and Jöckel, Karl-Heinz and Moebus, Susanne and Carr, J. Jeffrey and Erbel, Raimund and Burke, Gregory L.},
	month = oct,
	year = {2015},
	pmid = {26449133},
	pmcid = {PMC4603537},
	pages = {1643--1653},
}

@article{score2_working_group_and_esc_cardiovascular_risk_collaboration_score2_2021,
	title = {{SCORE2} risk prediction algorithms: new models to estimate 10-year risk of cardiovascular disease in {Europe}},
	volume = {42},
	issn = {0195-668X},
	shorttitle = {{SCORE2} risk prediction algorithms},
	url = {https://doi.org/10.1093/eurheartj/ehab309},
	doi = {10.1093/eurheartj/ehab309},
	number = {25},
	urldate = {2024-10-08},
	journal = {European Heart Journal},
	author = {{SCORE2 working group and ESC Cardiovascular risk collaboration}},
	month = jul,
	year = {2021},
	pages = {2439--2454},
}

@article{allen_automated_2024,
	title = {Automated {Breast} {Arterial} {Calcification} {Score} {Is} {Associated} {With} {Cardiovascular} {Outcomes} and {Mortality}},
	volume = {3},
	issn = {2772-963X},
	url = {https://www.sciencedirect.com/science/article/pii/S2772963X24005490},
	doi = {10.1016/j.jacadv.2024.101283},
	number = {11},
	urldate = {2024-10-04},
	journal = {JACC: Advances},
	author = {Allen, Tara Shrout and Bui, Quan M. and Petersen, Gregory M. and Mantey, Richard and Wang, Junhao and Nerlekar, Nitesh and Eghtedari, Mohammad and Daniels, Lori B.},
	month = nov,
	year = {2024},
	pages = {101283},
}

@article{jeong_emory_2023,
	title = {The {EMory} {BrEast} imaging {Dataset} ({EMBED}): {A} {Racially} {Diverse}, {Granular} {Dataset} of 3.4 {Million} {Screening} and {Diagnostic} {Mammographic} {Images}},
	volume = {5},
	issn = {2638-6100},
	shorttitle = {The {EMory} {BrEast} imaging {Dataset} ({EMBED})},
	url = {http://pubs.rsna.org/doi/10.1148/ryai.220047},
	doi = {10.1148/ryai.220047},
	language = {en},
	number = {1},
	urldate = {2024-09-30},
	journal = {Radiology: Artificial Intelligence},
	author = {Jeong, Jiwoong J. and Vey, Brianna L. and Bhimireddy, Ananth and Kim, Thomas and Santos, Thiago and Correa, Ramon and Dutt, Raman and Mosunjac, Marina and Oprea-Ilies, Gabriela and Smith, Geoffrey and Woo, Minjae and McAdams, Christopher R. and Newell, Mary S. and Banerjee, Imon and Gichoya, Judy and Trivedi, Hari},
	month = jan,
	year = {2023},
	pages = {e220047},
}

@article{guo_scu-net_2021,
	title = {{SCU}-{Net}: {A} deep learning method for segmentation and quantification of breast arterial calcifications on mammograms.},
	volume = {48},
	url = {http://dx.doi.org/10.1002/mp.15017},
	doi = {10.1002/mp.15017},
	number = {10},
	urldate = {2023-10-25},
	journal = {Medical Physics},
	author = {Guo, Xiaoyuan and O'Neill, W Charles and Vey, Brianna and Yang, Tianen Christopher and Kim, Thomas J and Ghassemi, Maryzeh and Pan, Ian and Gichoya, Judy Wawira and Trivedi, Hari and Banerjee, Imon},
	month = oct,
	year = {2021},
	pages = {5851--5861},
}

@article{mikhail_coronary_2005,
	title = {Coronary heart disease in women},
	volume = {331},
	copyright = {http://www.bmj.org/licenses/tdm/1.0/terms-and-conditions.html},
	issn = {0959-8138, 1468-5833},
	url = {https://www.bmj.com/lookup/doi/10.1136/bmj.331.7515.467},
	doi = {10.1136/bmj.331.7515.467},
	language = {en},
	number = {7515},
	urldate = {2024-12-11},
	journal = {BMJ},
	author = {Mikhail, Ghada W},
	month = sep,
	year = {2005},
	pages = {467--468},
}

@article{khan_development_2024,
	title = {Development and {Validation} of the {American} {Heart} {Association}’s {PREVENT} {Equations}},
	volume = {149},
	issn = {0009-7322, 1524-4539},
	url = {https://www.ahajournals.org/doi/10.1161/CIRCULATIONAHA.123.067626},
	doi = {10.1161/CIRCULATIONAHA.123.067626},
	language = {en},
	number = {6},
	urldate = {2024-12-11},
	journal = {Circulation},
	author = {Khan, Sadiya S. and Matsushita, Kunihiro and Sang, Yingying and Ballew, Shoshana H. and Grams, Morgan E. and Surapaneni, Aditya and Blaha, Michael J. and Carson, April P. and Chang, Alexander R. and Ciemins, Elizabeth and Go, Alan S. and Gutierrez, Orlando M. and Hwang, Shih-Jen and Jassal, Simerjot K. and Kovesdy, Csaba P. and Lloyd-Jones, Donald M. and Shlipak, Michael G. and Palaniappan, Latha P. and Sperling, Laurence and Virani, Salim S. and Tuttle, Katherine and Neeland, Ian J. and Chow, Sheryl L. and Rangaswami, Janani and Pencina, Michael J. and Ndumele, Chiadi E. and Coresh, Josef and {for the Chronic Kidney Disease Prognosis Consortium and the American Heart Association Cardiovascular-Kidney-Metabolic Science Advisory Group}},
	month = feb,
	year = {2024},
	pages = {430--449},
}

@article{duhn_breast_2011,
	title = {Breast {Arterial} {Calcification}: {A} {Marker} of {Medial} {Vascular} {Calcification} in {Chronic} {Kidney} {Disease}},
	volume = {6},
	issn = {1555-9041},
	shorttitle = {Breast {Arterial} {Calcification}},
	url = {https://journals.lww.com/01277230-201102000-00020},
	doi = {10.2215/CJN.07190810},
	language = {en},
	number = {2},
	urldate = {2024-12-11},
	journal = {Clinical Journal of the American Society of Nephrology},
	author = {Duhn, Valerie and D'Orsi, Ellen T. and Johnson, Samuel and D'Orsi, Carl J. and Adams, Amy L. and O'Neill, W. Charles},
	month = feb,
	year = {2011},
	pages = {377--382},
}

@article{kemmeren_arterial_1998,
	title = {Arterial {Calcification} {Found} on {Breast} {Cancer} {Screening} {Mammograms} and {Cardiovascular} {Mortality} in {Women}: {The} {DOM} {Project}},
	volume = {147},
	issn = {0002-9262, 1476-6256},
	shorttitle = {Arterial {Calcification} {Found} on {Breast} {Cancer} {Screening} {Mammograms} and {Cardiovascular} {Mortality} in {Women}},
	url = {https://academic.oup.com/aje/article-lookup/doi/10.1093/oxfordjournals.aje.a009455},
	doi = {10.1093/oxfordjournals.aje.a009455},
	language = {en},
	number = {4},
	urldate = {2024-12-11},
	journal = {American Journal of Epidemiology},
	author = {Kemmeren, J. M. and Van Noord, P. A. H. and Beijerinck, D. and Fracheboud, J. and Banga, J.-D. and Van Der Graaf, Y.},
	month = feb,
	year = {1998},
	pages = {333--341},
}

@article{iribarren_breast_2004,
	title = {Breast {Vascular} {Calcification} and {Risk} of {Coronary} {Heart} {Disease}, {Stroke}, and {Heart} {Failure}},
	volume = {13},
	copyright = {http://www.liebertpub.com/nv/resources-tools/text-and-data-mining-policy/121/},
	issn = {1540-9996, 1931-843X},
	url = {http://www.liebertpub.com/doi/10.1089/154099904323087060},
	doi = {10.1089/154099904323087060},
	language = {en},
	number = {4},
	urldate = {2024-12-11},
	journal = {Journal of Women's Health},
	author = {Iribarren, Carlos and Go, Alan S. and Tolstykh, Irina and Sidney, Stephen and Johnston, S. Claiborne and Spring, David B.},
	month = may,
	year = {2004},
	pages = {381--389},
}

@article{manzoor_progression_2018,
	title = {Progression of {Medial} {Arterial} {Calcification} in {CKD}},
	volume = {3},
	issn = {24680249},
	url = {https://linkinghub.elsevier.com/retrieve/pii/S2468024918301578},
	doi = {10.1016/j.ekir.2018.07.011},
	language = {en},
	number = {6},
	urldate = {2024-12-11},
	journal = {Kidney International Reports},
	author = {Manzoor, Shumila and Ahmed, Syed and Ali, Arshad and Han, Kum Hyun and Sechopoulos, Ioannis and O’Neill, Ansley and Fei, Baowei and O’Neill, W. Charles},
	month = nov,
	year = {2018},
	pages = {1328--1335},
}

@article{alappan_warfarin_2020,
	title = {Warfarin {Accelerates} {Medial} {Arterial} {Calcification} in {Humans}},
	volume = {40},
	issn = {1079-5642, 1524-4636},
	url = {https://www.ahajournals.org/doi/10.1161/ATVBAHA.119.313879},
	doi = {10.1161/ATVBAHA.119.313879},
	language = {en},
	number = {5},
	urldate = {2024-12-11},
	journal = {Arteriosclerosis, Thrombosis, and Vascular Biology},
	author = {Alappan, Harish R. and Kaur, Gurleen and Manzoor, Shumila and Navarrete, Jose and O’Neill, W. Charles},
	month = may,
	year = {2020},
	pages = {1413--1419},
}

@article{alappan_vascular_2020,
	title = {Vascular {Calcification} {Slows} {But} {Does} {Not} {Regress} {After} {Kidney} {Transplantation}},
	volume = {5},
	issn = {24680249},
	url = {https://linkinghub.elsevier.com/retrieve/pii/S2468024920316132},
	doi = {10.1016/j.ekir.2020.09.039},
	language = {en},
	number = {12},
	urldate = {2024-12-11},
	journal = {Kidney International Reports},
	author = {Alappan, Harish R. and Vasanth, Payaswini and Manzoor, Shumila and O’Neill, W. Charles},
	month = dec,
	year = {2020},
	pages = {2212--2217},
}

@inproceedings{urooj_multi-task_2024,
	address = {San Francisco},
	title = {A {Multi}-{Task} {Learning} {Approach} for {Segmentation} of {Breast} {Arterial} {Calcifications} in {Screening} {Mammograms}},
	booktitle = {{AMIA} 2024 {Annual} {Symposium}},
	publisher = {AMIA},
	author = {Urooj, Aisha and Dapamede, Theo and Patel, Bhavika and Ayoub, Chadi and Arsanjani, Reza and O'Neill, W. Charles and Trivedi, Hari and Banerjee, Imon},
	year = {2024},
}

@article{tsao_heart_2022,
	title = {Heart {Disease} and {Stroke} {Statistics}—2022 {Update}: {A} {Report} {From} the {American} {Heart} {Association}},
	volume = {145},
	issn = {0009-7322, 1524-4539},
	shorttitle = {Heart {Disease} and {Stroke} {Statistics}—2022 {Update}},
	url = {https://www.ahajournals.org/doi/10.1161/CIR.0000000000001052},
	doi = {10.1161/CIR.0000000000001052},
	language = {en},
	number = {8},
	urldate = {2024-12-11},
	journal = {Circulation},
	author = {Tsao, Connie W. and Aday, Aaron W. and Almarzooq, Zaid I. and Alonso, Alvaro and Beaton, Andrea Z. and Bittencourt, Marcio S. and Boehme, Amelia K. and Buxton, Alfred E. and Carson, April P. and Commodore-Mensah, Yvonne and Elkind, Mitchell S.V. and Evenson, Kelly R. and Eze-Nliam, Chete and Ferguson, Jane F. and Generoso, Giuliano and Ho, Jennifer E. and Kalani, Rizwan and Khan, Sadiya S. and Kissela, Brett M. and Knutson, Kristen L. and Levine, Deborah A. and Lewis, Tené T. and Liu, Junxiu and Loop, Matthew Shane and Ma, Jun and Mussolino, Michael E. and Navaneethan, Sankar D. and Perak, Amanda Marma and Poudel, Remy and Rezk-Hanna, Mary and Roth, Gregory A. and Schroeder, Emily B. and Shah, Svati H. and Thacker, Evan L. and VanWagner, Lisa B. and Virani, Salim S. and Voecks, Jenifer H. and Wang, Nae-Yuh and Yaffe, Kristine and Martin, Seth S. and {on behalf of the American Heart Association Council on Epidemiology and Prevention Statistics Committee and Stroke Statistics Subcommittee}},
	month = feb,
	year = {2022},
}

@article{bosco_major_2021,
	title = {Major adverse cardiovascular event definitions used in observational analysis of administrative databases: a systematic review},
	volume = {21},
	issn = {1471-2288},
	shorttitle = {Major adverse cardiovascular event definitions used in observational analysis of administrative databases},
	url = {https://bmcmedresmethodol.biomedcentral.com/articles/10.1186/s12874-021-01440-5},
	doi = {10.1186/s12874-021-01440-5},
	language = {en},
	number = {1},
	urldate = {2024-12-11},
	journal = {BMC Medical Research Methodology},
	author = {Bosco, Elliott and Hsueh, Leon and McConeghy, Kevin W. and Gravenstein, Stefan and Saade, Elie},
	month = dec,
	year = {2021},
	pages = {241},
}

@misc{american_college_of_cardiology_ascvd_nodate,
	title = {{ASCVD} {Risk} {Estimator} {Plus}},
	url = {https://tools.acc.org/ascvd-risk-estimator-plus/#!/calculate/estimate/},
	author = {American College of Cardiology},
}

@article{wenger_coronary_2003,
	series = {Heart {Disease} in {Women}, {Part} {I}},
	title = {Coronary heart disease: the female heart is vulnerable},
	volume = {46},
	issn = {0033-0620},
	shorttitle = {Coronary heart disease},
	url = {https://www.sciencedirect.com/science/article/pii/S0033062003001166},
	doi = {10.1016/j.pcad.2003.08.003},
	number = {3},
	urldate = {2024-12-18},
	journal = {Progress in Cardiovascular Diseases},
	author = {Wenger, Nanette K},
	month = nov,
	year = {2003},
	pages = {199--229},
}

@article{kelsey_results_1993,
	title = {Results of percutaneous transluminal coronary angioplasty in women. 1985-1986 {National} {Heart}, {Lung}, and {Blood} {Institute}'s {Coronary} {Angioplasty} {Registry}.},
	volume = {87},
	url = {https://www.ahajournals.org/doi/10.1161/01.CIR.87.3.720},
	doi = {10.1161/01.CIR.87.3.720},
	number = {3},
	urldate = {2024-12-18},
	journal = {Circulation},
	author = {Kelsey, S F and James, M and Holubkov, A L and Holubkov, R and Cowley, M J and Detre, K M},
	month = mar,
	year = {1993},
	note = {Publisher: American Heart Association},
	pages = {720--727},
}

@article{oneill_breast_2014,
	title = {Breast arterial calcification in chronic kidney disease: absence of smooth muscle apoptosis and osteogenic transdifferentiation},
	volume = {85},
	issn = {0085-2538},
	shorttitle = {Breast arterial calcification in chronic kidney disease},
	url = {https://www.sciencedirect.com/science/article/pii/S0085253815562248},
	doi = {10.1038/ki.2013.351},
	number = {3},
	urldate = {2025-01-09},
	journal = {Kidney International},
	author = {O'Neill, W. Charles and Adams, Amy L.},
	month = mar,
	year = {2014},
	pages = {668--676},
}

@article{borden_reducing_2022,
	title = {Reducing {Cardiovascular} {Risk} in the {Medicare} {Million} {Hearts} {Risk} {Reduction} {Model}: {Insights} {From} the {National} {Cardiovascular} {Data} {Registry} {PINNACLE} {Registry}},
	volume = {15},
	shorttitle = {Reducing {Cardiovascular} {Risk} in the {Medicare} {Million} {Hearts} {Risk} {Reduction} {Model}},
	url = {https://www.ahajournals.org/doi/10.1161/CIRCOUTCOMES.121.007908},
	doi = {10.1161/CIRCOUTCOMES.121.007908},
	number = {4},
	urldate = {2025-01-09},
	journal = {Circulation: Cardiovascular Quality and Outcomes},
	author = {Borden, William B. and Wang, Jingyan and Jones, Philip and Tang, Yuanyuan and Contreras, Johanna and Daugherty, Stacie L. and Desai, Nihar R. and Virani, Salim S. and Wasfy, Jason H. and Maddox, Thomas M.},
	month = apr,
	year = {2022},
	note = {Publisher: American Heart Association},
	pages = {e007908},
}

@article{arnett_2019_2019,
	title = {2019 {ACC}/{AHA} {Guideline} on the {Primary} {Prevention} of {Cardiovascular} {Disease}: {Executive} {Summary}: {A} {Report} of the {American} {College} of {Cardiology}/{American} {Heart} {Association} {Task} {Force} on {Clinical} {Practice} {Guidelines}},
	volume = {140},
	shorttitle = {2019 {ACC}/{AHA} {Guideline} on the {Primary} {Prevention} of {Cardiovascular} {Disease}},
	url = {https://www.ahajournals.org/doi/10.1161/CIR.0000000000000677},
	doi = {10.1161/CIR.0000000000000677},
	number = {11},
	urldate = {2025-01-22},
	journal = {Circulation},
	author = {Arnett, Donna K. and Blumenthal, Roger S. and Albert, Michelle A. and Buroker, Andrew B. and Goldberger, Zachary D. and Hahn, Ellen J. and Himmelfarb, Cheryl Dennison and Khera, Amit and Lloyd-Jones, Donald and McEvoy, J. William and Michos, Erin D. and Miedema, Michael D. and Muñoz, Daniel and Smith, Sidney C. and Virani, Salim S. and Williams, Kim A. and Yeboah, Joseph and Ziaeian, Boback},
	month = sep,
	year = {2019},
	note = {Publisher: American Heart Association},
	pages = {e563--e595},
}

@article{rotter_breast_2008,
	title = {Breast arterial calcifications ({BACs}) found on screening mammography and their association with cardiovascular disease},
	volume = {15},
	issn = {1072-3714},
	doi = {10.1097/gme.0b013e3181405d0a},
	language = {eng},
	number = {2},
	journal = {Menopause (New York, N.Y.)},
	author = {Rotter, Michelle A. and Schnatz, Peter F. and Currier, Allen A. and O'Sullivan, David M.},
	year = {2008},
	pmid = {17917612},
	pages = {276--281},
}

@article{bae_association_2013,
	title = {Association of breast arterial calcifications, metabolic syndrome, and the 10-year coronary heart disease risk: a cross-sectional case-control study},
	volume = {22},
	issn = {1931-843X},
	shorttitle = {Association of breast arterial calcifications, metabolic syndrome, and the 10-year coronary heart disease risk},
	doi = {10.1089/jwh.2012.4148},
	language = {eng},
	number = {7},
	journal = {Journal of Women's Health (2002)},
	author = {Bae, Mi Jin and Lee, Sang Yeoup and Kim, Yun Jin and Lee, Jeong Gyu and Jeong, Dong Wook and Yi, Yu Hyeon and Cho, Young Hye and Choi, Eun Jung and Choo, Ki Seok},
	month = jul,
	year = {2013},
	pmid = {23790228},
	pages = {625--630},
}

@article{hendriks_breast_2015,
	title = {Breast arterial calcifications: a systematic review and meta-analysis of their determinants and their association with cardiovascular events},
	volume = {239},
	issn = {1879-1484},
	shorttitle = {Breast arterial calcifications},
	doi = {10.1016/j.atherosclerosis.2014.12.035},
	language = {eng},
	number = {1},
	journal = {Atherosclerosis},
	author = {Hendriks, Eva J. E. and de Jong, Pim A. and van der Graaf, Yolanda and Mali, Willem P. Th M. and van der Schouw, Yvonne T. and Beulens, Joline W. J.},
	month = mar,
	year = {2015},
	pmid = {25568948},
	pages = {11--20},
}

@article{lanzer_medial_2014,
	title = {Medial vascular calcification revisited: review and perspectives},
	volume = {35},
	issn = {0195-668X},
	shorttitle = {Medial vascular calcification revisited},
	url = {https://doi.org/10.1093/eurheartj/ehu163},
	doi = {10.1093/eurheartj/ehu163},
	number = {23},
	urldate = {2025-01-22},
	journal = {European Heart Journal},
	author = {Lanzer, Peter and Boehm, Manfred and Sorribas, Victor and Thiriet, Marc and Janzen, Jan and Zeller, Thomas and St Hilaire, Cynthia and Shanahan, Catherine},
	month = jun,
	year = {2014},
	pages = {1515--1525},
}

@article{schnatz_association_2011,
	title = {The association of breast arterial calcification and coronary heart disease},
	volume = {117},
	issn = {1873-233X},
	doi = {10.1097/AOG.0b013e318206c8cb},
	language = {eng},
	number = {2 Pt 1},
	journal = {Obstetrics and Gynecology},
	author = {Schnatz, Peter F. and Marakovits, Kimberly A. and O'Sullivan, David M.},
	month = feb,
	year = {2011},
	pmid = {21252734},
	pages = {233--241},
}

@article{lloyd-jones_use_2019,
	title = {Use of {Risk} {Assessment} {Tools} to {Guide} {Decision}-{Making} in the {Primary} {Prevention} of {Atherosclerotic} {Cardiovascular} {Disease}: {A} {Special} {Report} {From} the {American} {Heart} {Association} and {American} {College} of {Cardiology}},
	volume = {139},
	issn = {0009-7322, 1524-4539},
	shorttitle = {Use of {Risk} {Assessment} {Tools} to {Guide} {Decision}-{Making} in the {Primary} {Prevention} of {Atherosclerotic} {Cardiovascular} {Disease}},
	url = {https://www.ahajournals.org/doi/10.1161/CIR.0000000000000638},
	doi = {10.1161/CIR.0000000000000638},
	language = {en},
	number = {25},
	urldate = {2025-02-19},
	journal = {Circulation},
	author = {Lloyd-Jones, Donald M. and Braun, Lynne T. and Ndumele, Chiadi E. and Smith, Sidney C. and Sperling, Laurence S. and Virani, Salim S. and Blumenthal, Roger S.},
	month = jun,
	year = {2019},
}

@misc{american_cancer_society_cancer_2024,
	title = {Cancer {Prevention} \& {Early} {Detection}},
	url = {https://www.cancer.org/research/cancer-facts-statistics/cancer-prevention-early-detection.html},
	language = {en},
	urldate = {2025-02-19},
	author = {American Cancer Society},
	year = {2024},
}

@inproceedings{urooj_multi-task_2023,
	address = {New Orleans},
	title = {Multi-{Task} {Learning} for {Segmentation} of {Breast} {Arterial} {Calcifications} in {Mammograms}},
	booktitle = {Medical {Imaging} meets {NeurIPS}},
	author = {Urooj, Aisha and O'Neill, W. Charles and Trivedi, Hari and Banerjee, Imon},
	year = {2023},
}

@article{richman_adoption_2019,
	title = {Adoption of {Digital} {Breast} {Tomosynthesis} in {Clinical} {Practice}},
	volume = {179},
	issn = {2168-6106},
	url = {https://doi.org/10.1001/jamainternmed.2019.1058},
	doi = {10.1001/jamainternmed.2019.1058},
	number = {9},
	urldate = {2025-02-21},
	journal = {JAMA Internal Medicine},
	author = {Richman, Ilana B. and Hoag, Jessica R. and Xu, Xiao and Forman, Howard P. and Hooley, Regina and Busch, Susan H. and Gross, Cary P.},
	month = sep,
	year = {2019},
	pages = {1292--1295},
}

@article{lee_effect_2018,
	title = {The {Effect} of {Digital} {Breast} {Tomosynthesis} {Adoption} on {Facility}-{Level} {Breast} {Cancer} {Screening} {Volume}},
	volume = {211},
	issn = {0361-803X},
	url = {https://www.ncbi.nlm.nih.gov/pmc/articles/PMC6438161/},
	doi = {10.2214/AJR.17.19350},
	number = {5},
	urldate = {2025-02-21},
	journal = {AJR. American journal of roentgenology},
	author = {Lee, Christoph I. and Zhu, Weiwei and Onega, Tracy L. and Germino, Jessica and O’Meara, Ellen S. and Lehman, Constance D. and Henderson, Louise M. and Haas, Jennifer S. and Kerlikowske, Karla and Sprague, Brian L. and Rauscher, Garth H. and Tosteson, Anna N.A. and Alford-Teaster, Jennifer and Wernli, Karen J. and Miglioretti, Diana L.},
	month = nov,
	year = {2018},
	pmid = {30235000},
	pmcid = {PMC6438161},
	pages = {957--963},
}

@article{iribarren_breast_2022,
	title = {Breast {Arterial} {Calcification}: a {Novel} {Cardiovascular} {Risk} {Enhancer} among {Postmenopausal} {Women}},
	volume = {15},
	issn = {1941-9651},
	shorttitle = {Breast {Arterial} {Calcification}},
	url = {https://www.ncbi.nlm.nih.gov/pmc/articles/PMC8931858/},
	doi = {10.1161/CIRCIMAGING.121.013526},
	number = {3},
	urldate = {2025-02-21},
	journal = {Circulation. Cardiovascular imaging},
	author = {Iribarren, Carlos and Chandra, Malini and Lee, Catherine and Sanchez, Gabriela and Sam, Danny L. and Azamian, Farima Faith and Cho, Hyo-Min and Ding, Huanjun and Wong, Nathan D. and Molloi, Sabee},
	month = mar,
	year = {2022},
	pmid = {35290077},
	pmcid = {PMC8931858},
	pages = {e013526},
}

@article{yurdaisik_evaluation_2020,
	title = {Evaluation of the {Correlation} {Between} {Breast} {Artery} {Calcification} and {Coronary} {Artery} {Calcium} {Scores} in {Predicting} the {Risk} for {Cardiovascular} {Disease}},
	volume = {19},
	issn = {2149-5807, 2149-6048},
	url = {https://eajem.com/articles/doi/eajem.galenos.2020.16779},
	doi = {10.4274/eajem.galenos.2020.16779},
	number = {3},
	urldate = {2025-03-12},
	journal = {Eurasian Journal of Emergency Medicine},
	author = {Yurdaışık, Işıl and Nurili, Fuad},
	month = sep,
	year = {2020},
	pages = {136--141},
}

@article{ryan_breast_2017,
	title = {Breast arterial calcification association with coronary artery calcium scoring and implications for cardiovascular risk assessment in women},
	volume = {40},
	copyright = {http://onlinelibrary.wiley.com/termsAndConditions\#vor},
	issn = {0160-9289, 1932-8737},
	url = {https://onlinelibrary.wiley.com/doi/10.1002/clc.22702},
	doi = {10.1002/clc.22702},
	language = {en},
	number = {9},
	urldate = {2025-03-12},
	journal = {Clinical Cardiology},
	author = {Ryan, Angela J. and Choi, Andrew D. and Choi, Brian G. and Lewis, Jannet F.},
	month = sep,
	year = {2017},
	pages = {648--653},
}
\clearpage
\appendix
\section{Appendix}

\setcounter{figure}{0}
\renewcommand{\thefigure}{\Alph{section}\arabic{figure}}
\setcounter{table}{0}
\renewcommand{\thetable}{\Alph{section}\arabic{table}}

\subsection{ICD Codes for MACE criteria}

\begin{table}[htb]
\centering
\caption{ICD Codes for the MACE criteria}
\label{tab:icd}
\resizebox{0.95\textwidth}{!}{%
\begin{tabular}{lll}
\hline
\textbf{Outcome} & \textbf{ICD-9} & \textbf{ICD-10} \\
\hline
Non-fatal myocardial infarction & 410.0, 410.00, 410.01, 410.1, 410.10, 410.11, & I21.0, I21.01, I21.02, I21.09, I21.1, I21.11, \\
 & 410.20, 410.2, 410.21, 410.3, 410.30, 410.31, & I21.19, I21.2, I21.21, I21.29, I21.3, I21.4, \\
 & 410.40, 410.4, 410.41, 410.5, 410.50, 410.51, & I21.9 \\
 & 410.6, 410.60, 410.61, 410.7, 410.70, 410.71, & \\
 & 410.8, 410.80, 410.81, 410.9, 410.90, 410.91 & \\
\hline
Non-fatal ischemic stroke & 433.01, 433.11, 433.21, 433.31, 433.81, & I63.* \\
 & 433.91, 434.01, 434.11, 434.91 & \\
\hline
Acute Heart Failure & 428.43, 428.33, 428.23, 428.41 & I50.813, I50.43, I50.33, I50.23, I50.21 \\
\hline
\end{tabular}}
\end{table}

Pre-selected ICD codes extracted from EHR \cite{bosco_major_2021}

\subsection{Patient Flow Diagram}

\begin{figure}[htbp]
\centering
\includegraphics[width=\linewidth]{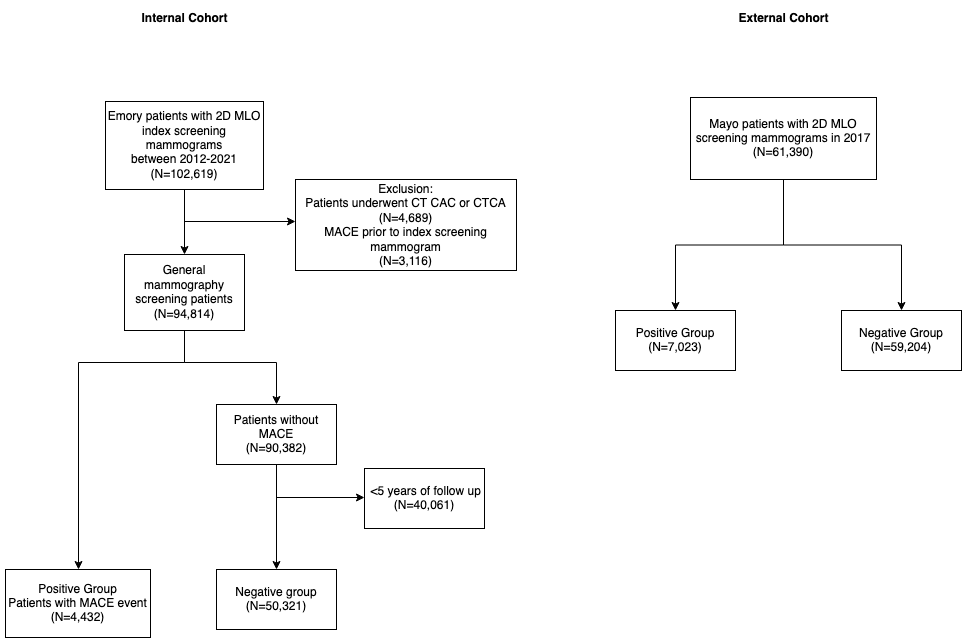}
\caption{Patient selection flow diagram for internal and external cohort.}
\label{fig:patient_flow}
\end{figure}

\subsection{BAC Severity Classification}

\begin{figure}[htbp]
\centering
\includegraphics[width=\linewidth]{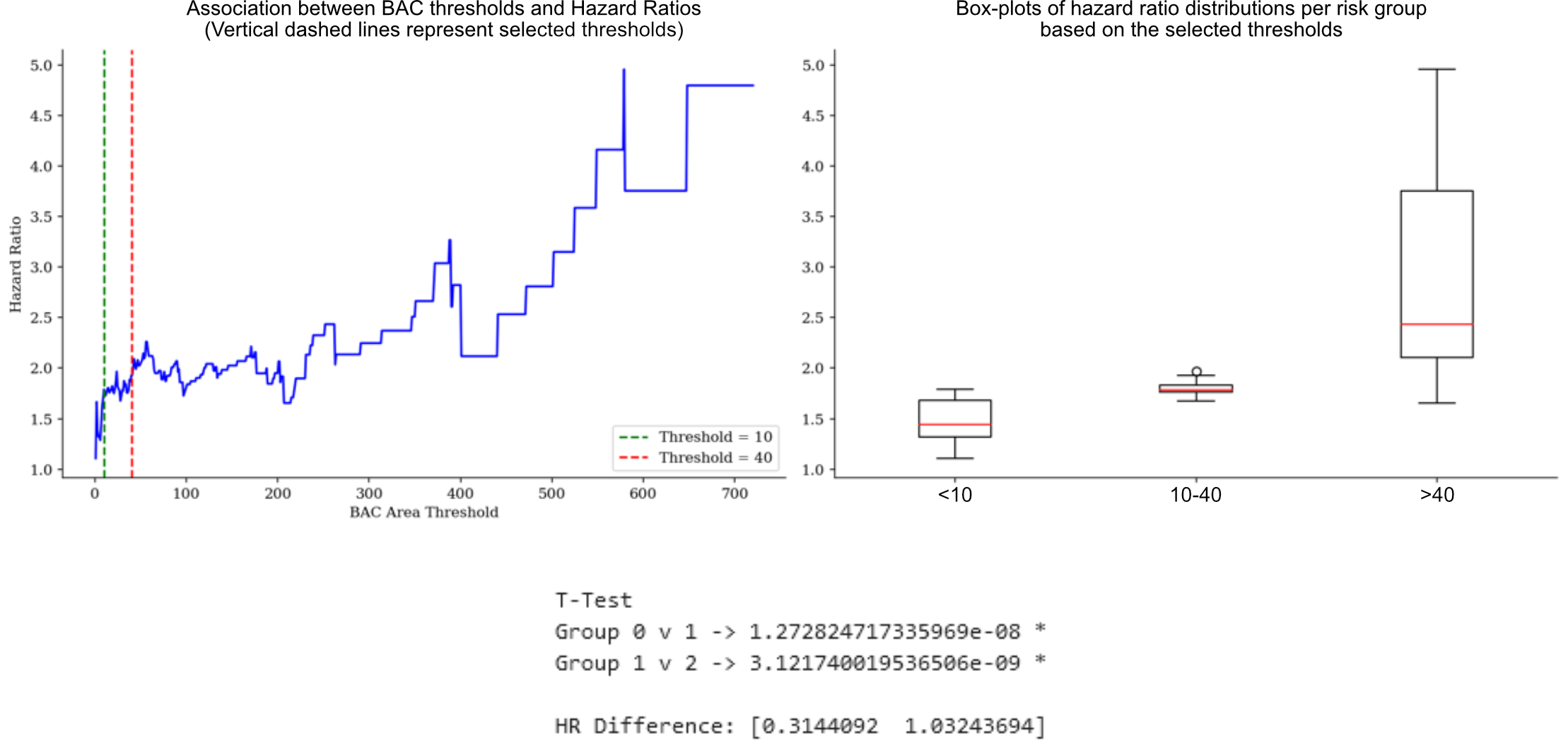}
\caption{Patient selection flow diagram for internal and external cohort.}
\label{fig:bac_empirical_severity_classification}
\end{figure}

To determine optimal breast arterial calcification area (BAC) thresholds for severity classification, we performed a threshold analysis using univariate Cox proportional hazards models. We systematically evaluated different BAC cutoff values at 5mm² intervals between 5-700mm². The thresholds of 10mm² and 40mm² were selected as they maximized the hazard ratio differences between adjacent severity groups while maintaining adequate sample sizes in each group. After establishing these thresholds, we validated the groupings by comparing clinical outcomes between mild (<10mm²), moderate (10-40mm²), and severe (>40mm²) groups using independent samples t-tests.

\subsection{Sensitivity Analysis to Validate Calculated ASCVD Scores}

\begin{figure}[htbp]
\centering
\includegraphics[width=\linewidth]{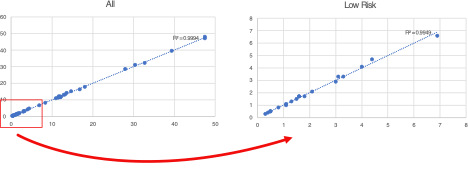}
\caption{Sensitivity Analysis to Validate Calculated ASCVD Scores.}
\label{fig:ascvd_validate}
\end{figure}

Calibration analysis was performed to assess the relationship between ASCVD risk scores calculated by our automated script developed from the ASCVD equation and the ground truth ASCVD risk scores calculated from the online tool \cite{american_college_of_cardiology_ascvd_nodate}.

\subsection{Presence of mammographic artifacts}

\begin{figure}[htbp]
\centering
\includegraphics[width=0.95\linewidth]{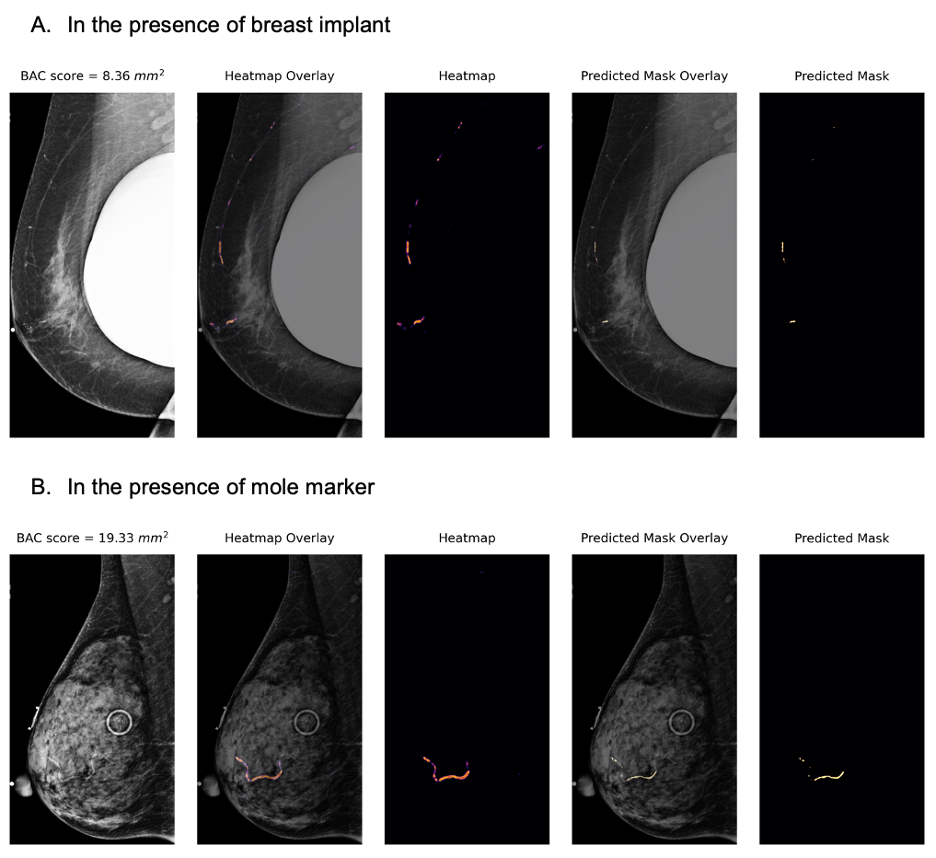}
\caption{Examples of the BAC segmentation in the presence of mammographic artifacts.}
\label{fig:artifacts}
\end{figure}

\subsection{Examples of False Positives}

\begin{figure}[htbp]
\centering
\includegraphics[width=0.95\linewidth]{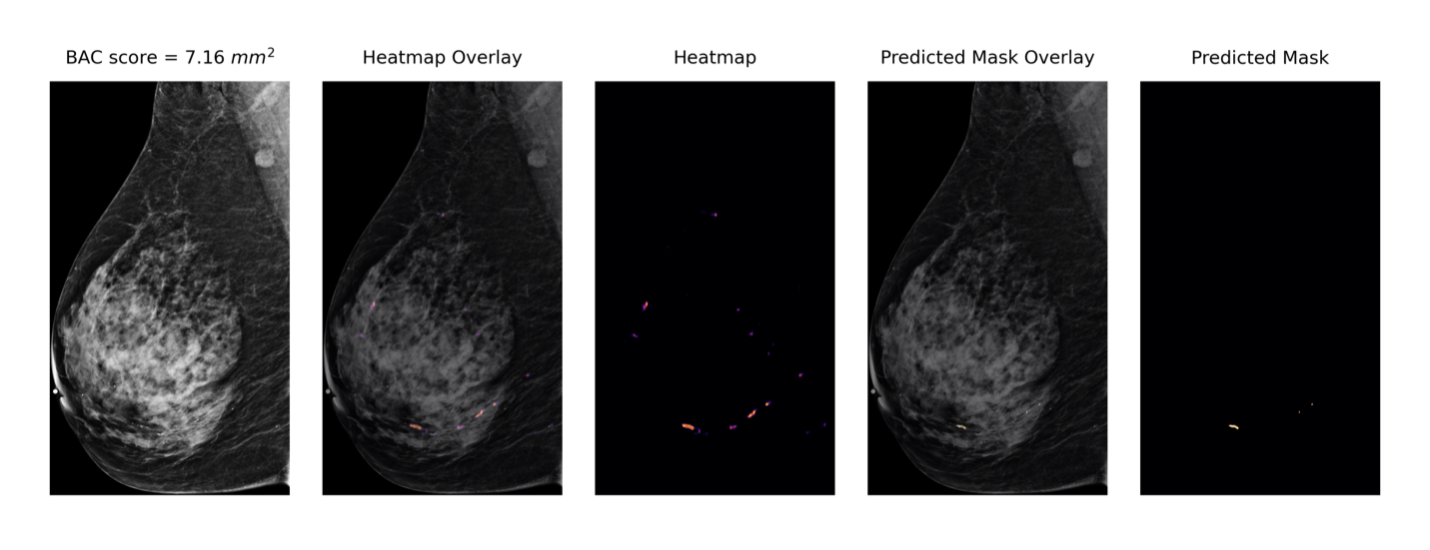}
\caption{Examples of false positives from non-vascular calcifications.}
\label{fig:false_positives}
\end{figure}
\end{document}